\documentclass[lettersize,journal]{IEEEtran}
\usepackage{amsmath,amsfonts}
\usepackage{algorithm}
\usepackage{algorithmic}
\usepackage{array}
\usepackage[caption=false,font=normalsize,labelfont=sf,textfont=sf]{subfig}
\usepackage{textcomp}
\usepackage{stfloats}
\usepackage{url}
\usepackage{verbatim}
\usepackage{booktabs}
\usepackage{xcolor}
\usepackage{subcaption}
\usepackage{makecell}
\usepackage{hyperref}
\usepackage{graphicx}
\usepackage{multirow}
\usepackage{cite}
\begin{document}

\title{Proxy Model-Guided Reinforcement Learning for Client Selection in Federated Recommendation}

\author{
    Liang Qu$^{1}$~\IEEEmembership{Member,~IEEE},
    Jianxin Li$^{1\ast}$~\IEEEmembership{Senior Member,~IEEE},
    Wei Yuan$^{2}$,
    Penghui Ruan$^{3}$,\\
    Yuhui Shi$^{3}$~\IEEEmembership{Fellow,~IEEE},
    Hongzhi Yin$^{2\ast}$~\IEEEmembership{Senior Member,~IEEE}%
    \thanks{$^{1}$School of Business and Law, Edith Cowan University, Perth, Australia.}%
    \thanks{$^{2}$School of Electrical Engineering and Computer Science, The University of Queensland, Brisbane, Australia.}%
    \thanks{$^{3}$Department of Computer Science and Engineering, Southern University of Science and Technology, Shenzhen, China.}%
    \thanks{$^{\ast}$Corresponding authors: Jianxin Li (jianxin.li@ecu.edu.au) and Hongzhi Yin (h.yin1@uq.edu.au).}%
}

\markboth{Journal of \LaTeX\ Class Files,~Vol.~14, No.~8, August~2021}%
{Shell \MakeLowercase{\textit{et al.}}: A Sample Article Using IEEEtran.cls for IEEE Journals}


\maketitle

\begin{abstract}
Federated recommender systems (FedRSs) have emerged as a promising privacy-preserving paradigm, enabling personalized recommendation services without exposing users’ raw data. By keeping data local and relying on a central server to coordinate training across distributed clients, FedRSs protect user privacy while collaboratively learning global models.
However, most existing FedRS frameworks adopt fully random client selection strategy in each training round, overlooking the statistical heterogeneity of user data arising from diverse preferences and behavior patterns, thereby resulting in suboptimal model performance.
While some client selection strategies have been proposed in the broader federated learning (FL) literature, these methods are typically designed for generic tasks and fail to address the unique challenges of recommendation scenarios, such as expensive contribution evaluation due to the large number of clients, and sparse updates resulting from long-tail item distributions.
To bridge this gap, we propose ProxyRL-FRS, a proxy model-guided reinforcement learning framework tailored for client selection in federated recommendation. 
Specifically, we first introduce ProxyNCF, a dual-branch model deployed on each client, which augments standard Neural Collaborative Filtering with an additional proxy model branch that provides lightweight contribution estimation, thus eliminating the need for expensive per-round local training traditionally required to evaluate a client’s contribution.
Furthermore, we design a staleness-aware (SA) reinforcement learning agent that selects clients based on the proxy-estimated contribution, and is guided by a reward function balancing recommendation accuracy and embedding staleness, thereby enriching the update coverage of item embeddings.
Experiments conducted on three public recommendation datasets demonstrate that ProxyRL-FRS outperforms state-of-the-art baselines, achieving faster convergence and improved recommendation accuracy.

\end{abstract}


\section{Introduction}
Recommender systems (RSs) have become an essential component of many real-world applications, driving services such as video recommendations on YouTube \cite{covington2016deep} and product suggestions on Amazon \cite{linden2003amazon}.
These applications deliver personalized content by learning user-item interaction data, such as likes or purchases, to infer individual user preferences. However, RSs generally require users to share sensitive user-item interaction data with application platforms for the purpose of capturing user preferences, raising significant privacy 
concerns. In light of these concerns, many privacy regulations, such as the General Data Protection Regulation (GDPR\footnote{https://gdpr-info.eu/}), have been introduced in recent years. Consequently, there is an urgent need to develop solutions that strike a balance between preserving user privacy and maintaining the effectiveness of recommendation services.

\begin{figure}
    \centering
    \includegraphics[width=1\linewidth]{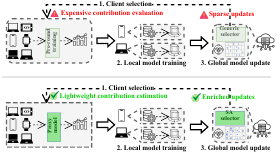}
    \caption{Comparison between client selection methods designed for generic FL tasks (top) and our proposed method tailored for federated recommendation (bottom).}
    \label{fig:FLvsFRS}
\end{figure}

Inspired by the promising advancements of federated learning in privacy preserving machine learning \cite{zhang2021survey}, recent studies have explored its application in recommender systems, leading to the development of federated recommender systems (FedRSs) \cite{sun2022survey,yang2020federated,alamgir2022federated,asad2023comprehensive,harasic2024recent,yang2025survey,10.1145/3708982}. FedRSs aim to address privacy concerns in recommendation tasks by ensuring that users' private data, such as user-item interaction data, remain strictly on their local devices, while leveraging a central server to coordinate training a global model across distributed clients. Typically, as illustrated in Figure \ref{fig:FLvsFRS}, a FedRS training pipeline consists of three sequential steps: (1) client selection, where a central server selects a subset of clients to participate in the current round of training based on a client selection strategy; (2) local model training, where the selected clients update their local models using local private data; and (3) global model update, where the server aggregates the local model knowledge (e.g., model gradients/parameters) to update the global model. As the first step in the FedRS training pipeline, client selection determines which user data contributes to model training, thereby laying the foundation for the subsequent local model training and global model update steps. However, most existing FedRSs adopt a fully random client selection strategy in each training round \cite{yin2024device}, treating all clients as equally informative and assigning them equal probabilities of selection. However, this assumption does not hold in real-world scenarios, as it overlooks the inherent statistical heterogeneity of user data, which results from the diverse preferences and behaviors of different users. Consequently, the data distributed across clients is often non-independent and non-identically distributed (non-IID), making random client selection suboptimal for effective model training.

In response to this challenge, a number of research in federated learning has explored non-random client selection strategies \cite{fu2023client, mayhoub2024review, li2024comprehensive}, which can be broadly categorized into three main types: (1) Heuristic-based methods. These methods select clients based on predefined criteria for client contribution, such as local data quality \cite{deng2021auction}, gradient norms \cite{marnissi2024client}, clients’ Shapley values \cite{liu2022gtg,jia2019towards}, or local training loss \cite{cho2022towards}. While straightforward, heuristic methods typically depend on expert-designed rules, making them difficult to generalize across different tasks and limiting their adaptability. (2) Clustering-based methods. These methods group similar clients using metadata or model parameter distributions \cite{muhammad2020fedfast, imran2023refrs, luo2022personalized}, and then select clients from different clusters to enhance selection diversity. However, they often resort to random sampling within each cluster, which does not fully resolve the issue of suboptimal selection. Moreover, their performance is highly sensitive to the number of clusters, which is often difficult to determine in practice. (3) Learning-based methods. These methods formulate client selection as a sequential decision-making problem, often modeled as a Markov decision process. They typically employ multi-armed bandit algorithms \cite{cho2020bandit, ko2021joint} or reinforcement learning techniques \cite{deng2021auction, albelaihi2023deep,zhang2022multi,wang2025reinforcement} to select clients by maximizing a reward function that reflects desired outcomes, generally offering better generalization and adaptability across different settings.

\begin{figure*}[t]
\centering
\includegraphics[width=1\textwidth]{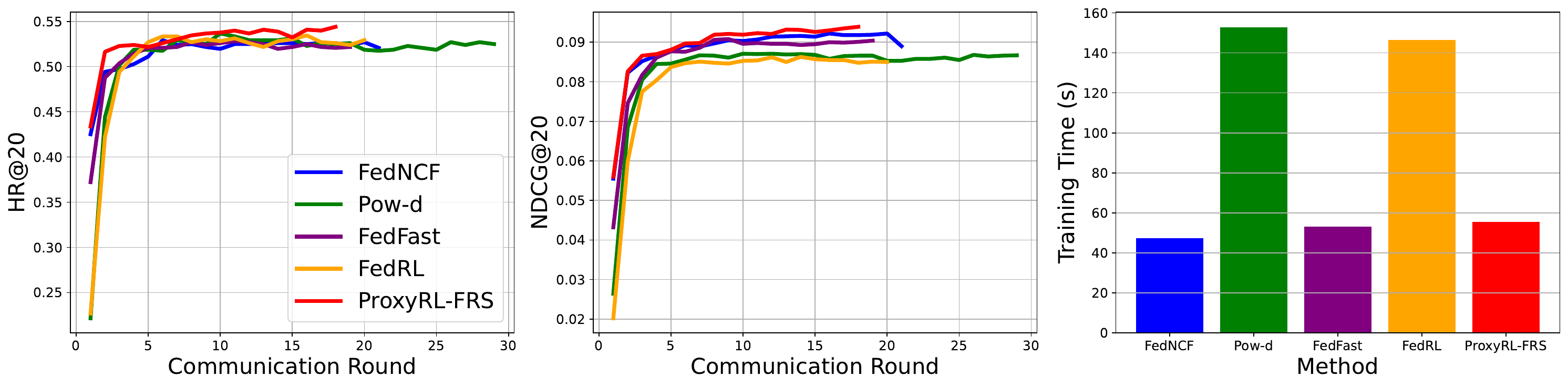} 
\caption{Comparison of five client selection strategies on the MovieLens-100k \cite{harper2015movielens}: (1) Random (FedNCF \cite{perifanis2022federated}), (2) Heuristic (Pow-d \cite{cho2022towards}), (3) Clustering-based (FedFast \cite{muhammad2020fedfast}), (4) Learning-based (FedRL \cite{di2024fedrl}), and (5) Proposed method (ProxyRL-FRS). Evaluation is based on HR@20, NDCG@20, and cumulative training time.}
\label{fig:PreliminaryExperiment}
\end{figure*}

While the above client selection methods have achieved promising results, most are designed for generic tasks such as computer vision and natural language processing \cite{zhang2022multi, li2024comprehensive}, and are hence referred to as \textit{generic selectors} in this paper.
Directly adapting these generic selectors to recommendation tasks is often non-trivial, as shown in the top of Figure~\ref{fig:FLvsFRS}.
These methods often overlook the unique challenges inherent in recommendation settings, including:
\textbf{(1) Expensive contribution evaluation:}
Unlike generic FL scenarios involving institutional clients (e.g., hospitals or banks), recommendation tasks involve a substantially larger number of clients, typically individual user devices (e.g., smartphones or laptops). Consequently, applying generic selectors is often highly time-consuming and impractical, as they typically require candidate clients to perform pre-round local training to evaluate their contributions. For instance, heuristic methods (e.g., Pow-d \cite{cho2022towards}) prioritize clients with higher local loss, while reinforcement learning-based methods (e.g., FedRL \cite{di2024fedrl}) leverage per-round local training signals to construct the environment state. As empirically shown by the preliminary results in Figure~\ref{fig:PreliminaryExperiment}, these methods incur significant pre-round training time due to the large number of candidate clients involved.
\textbf{(2) Sparse updates:}
User-item interaction data in recommendation tasks typically follows a long-tail distribution, resulting in inherently sparse local datasets. Unlike generic FL tasks, where each client generally has access to richer datasets and all model parameters (e.g., neural network weights) are updated in every training round, by contrast, FedRS models are dominated by large item embedding tables. Crucially, only the embeddings of items explicitly interacted with by users are updated during local training. This results in sparse global model updates, especially for long-tail items, thereby limiting the effectiveness of aggregation and resulting in suboptimal performance for generic selectors, as observed in Figure~\ref{fig:PreliminaryExperiment}.

To address the aforementioned challenges, we propose ProxyRL-FRS, a \underline{Proxy} model-guided \underline{R}einforcement \underline{L}earning framework tailored for \underline{F}ederated \underline{R}ecommender \underline{S}ystems. As illustrated in the bottom part of Figure~\ref{fig:FLvsFRS}, we first introduce ProxyNCF, a dual-branch model deployed on each client. In addition to the standard Neural Collaborative Filtering (NCF) \cite{he2017neural} branch used for local recommendation learning, we augment it with a lightweight proxy model branch that estimates the client’s contributions (e.g., training loss) without requiring full local training. This design eliminates the overhead of costly per-round training traditionally required for client evaluation.
Furthermore, we design a staleness-aware (SA) reinforcement learning agent that selects a subset of clients for each communication round based on proxy-estimated contributions. The reward function is carefully crafted to balance two objectives: maximizing recommendation accuracy and minimizing embedding staleness, thereby encouraging frequent updates to stale (e.g., long-tail) item embeddings.

In summary, the contributions of this work include the following three points:
\begin{itemize}
    \item We identify two key challenges in federated recommendation that are not well addressed by existing client selection methods: (1) the high cost of contribution evaluation due to the large number of clients, and (2) sparse global model updates resulting from long-tail item distributions.
    \item We propose ProxyRL-FRS, a proxy model-guided reinforcement learning framework tailored for federated recommendation. It integrates a lightweight dual-branch model (ProxyNCF) for efficient contribution estimation and a staleness-aware reinforcement learning agent that promotes both accuracy and embedding freshness.
    \item Extensive experiments on public recommendation datasets demonstrate that our method consistently achieves higher recommendation accuracy than baseline methods while maintaining comparable training efficiency.
\end{itemize}

The remainder of the paper is organized as follows: Section 2 reviews three categories of methods relevant to this work. Section 3 introduces the preliminaries and problem formulation. Section 4 details the proposed method. The experimental results are discussed in Section 5, followed by the conclusion in Section 6.

\section{Related Work}
In this section, we will introduce three categories of methods related to this work, including federated recommendation methods and client selection methods in generic FL tasks and federated recommendation.

\subsection{Federated Recommender Systems}
Federated recommender systems (FedRSs) \cite{li2024personalized,10.1145/3637528.3671702,zhang2024federated,zheng2025deckg,yuan2025robust,yuan2025ptf,yang2024pdc,yuan2024robust,qu2024towards,qu2023semi,yuan2024hide,yuan2024hetefedrec} aim to address privacy concerns by storing users' private data locally and utilizing a central server to collaborate with a group of clients to learn a global model. 
Generally, a typical federated recommender system (FedRS) pipeline consists of three key steps: client selection, local model training, and global model update. However, most current research on FedRSs primarily focuses on the latter two steps. For instance, in the local training step, various recommendation models have been employed, such as the pioneering matrix factorization-based method FCF \cite{ammad2019federated}, the deep neural network-based method FedNCF \cite{perifanis2022federated}, and the recent GNN-based method FedPerGNN \cite{wu2021fedgnn}. In the global model updating step, FedDSR \cite{huang2021feddsr} introduces a similarity aggregation strategy that aggregates only the gradients from similar clients based on their uploaded gradient information. Similarly, ReFRS \cite{imran2023refrs} proposes aggregating clients with semantic similarity to improve privacy and model performance.

However, client selection, which serves as the foundation for the latter two steps, has been largely overlooked in FedRS research. Most FedRSs adopt an unbiased client selection method, such as the commonly used random client selection approach (e.g., FedAvg \cite{mcmahan2017communication}), which assumes all clients are equivalent, resulting in suboptimal results.

\subsection{Client Selection in Federated Learning}
To address the challenges caused by statistical heterogeneity in local data, researchers have proposed various client selection strategies \cite{fu2023client, mayhoub2024review, li2024comprehensive}, which can be broadly divided into three categories:
(a) Heuristic Methods: These methods select clients based on predefined criteria. For example, some methods prioritize clients based on the quality of their local data \cite{deng2021auction}, while others utilize the norm of gradients after local training \cite{marnissi2024client}. Additionally, some approaches select clients based on their local training loss \cite{cho2022towards}, prioritizing clients with higher losses as their inclusion can lead to more significant improvements in the global model. However, heuristic methods often rely on a single predefined criterion, making them difficult to generalize across different problems, thus limiting their applicability and robustness.
(b) Clustering Methods: This type of approach seeks to increase the diversity of selected clients by grouping similar clients based on metadata or parameter distributions \cite{muhammad2020fedfast, imran2023refrs, luo2022personalized}. Once grouped, clients are sampled from each cluster to ensure balanced representation. Despite their potential to capture client diversity, these methods often default to random selection within clusters, which undermines their effectiveness. Moreover, the performance of these approaches is highly sensitive to the number of clusters, reducing their robustness in dynamic environments.
(c) Learning Methods: These methods leverage advanced algorithms to dynamically adjust client selection. By framing the problem as a decision-making task, approaches such as multi-armed bandits \cite{cho2020bandit, ko2021joint} or reinforcement learning \cite{deng2021auction, albelaihi2023deep,wang2025reinforcement,zhang2022multi} are used to maximize a reward function that reflects the goals of the selection process, such as optimizing accuracy or reducing communication overhead. These methods offer greater flexibility by adapting to the current state of the system, making them more versatile for federated learning.

However, these client selection methods primarily focus on non-recommendation tasks, where clients are often small-scale servers or data centers. In contrast, in the recommendation context, each client typically corresponds to a user's device, such as a smartphone, leading to a much larger number of clients compared to non-recommendation scenarios. This scale poses significant challenges when directly applying existing methods. For example, in heuristic methods and learning methods, requiring every client to perform local training in each round to obtain heuristic information such as local loss or gradient norms is extremely time-consuming. 

\subsection{Client Selection in Federated Recommendation}
In the domain of federated recommendation, client selection remains relatively underexplored, and only a few works attempt to address this challenge. For example, FedFast \cite{muhammad2020fedfast} is an early work that clusters clients into multiple groups based on their meta information, and then randomly selects clients within each cluster.
FedGST \cite{tang2024fedgst} proposes to use an influence function to assess the contributions of candidate clients, and clients with high influence scores are selected to participate in the next federated training round. FedRL \cite{di2024fedrl} proposes to employ a reinforcement learning agent to select clients based on indicators such as training loss, and utilizes a hypernetwork to generate personalized model parameters for local devices. FedACS \cite{xu2025fedacs} first formulates federated recommendation as a federated graph learning problem by partitioning the user-item interaction data into multiple user-item subgraphs. It then employs a multi-armed bandit algorithm to perform adaptive client selection. 

However, most existing methods rely on expensive contribution evaluation techniques, such as local training loss or influence estimation, which require clients to perform pre-round local training prior to selection, resulting in considerable computation overhead. Moreover, few studies explicitly consider the long-tail item distribution that is intrinsic to recommendation tasks, which leads to sparse updates and stale item embeddings if not properly addressed.

\section{Preliminary}
Before introducing our proposed method, we first outline the typical pipeline of federated recommender systems and then provide a formal definition of the client selection problem within the context of FedRSs. 

\begin{table}[h!]
\caption{Frequently used notations and descriptions}
\centering
\begin{tabular}{@{}c l@{}}
\toprule[1.5pt] 
\textbf{Notations} & \textbf{Descriptions} \\ 
\midrule[1pt] 
$\mathcal{U}$               & The set of users.       \\ 
$\mathcal{V}$               & The set of items.       \\ 
$u$                         & The user/client $u$.       \\
$v$                         & The item $v$. \\
$\mathcal{D}_{u}$           & The local user-item interaction dataset of user $u$. \\
$\mathcal{V}_{u}$           & The set of items that the user $u$ has interacted with. \\
$f_{u}(\cdot)$ & The local model of the user $u$. \\
$\textbf{u}$ & The user embedding of the user $u$. \\
$\textbf{v}$ & The item embedding of the item $v$. \\
$P_{u}$ & The local item embedding table of the user $u$. \\
$\Theta_{u}$ & The local model parameter of the user $u$. \\
$f_{s}(\cdot)$ & The global model on the server side. \\
$P_{s}$ & The global item embedding table. \\
$\Theta_{s}$ & The global model parameters. \\
$\mathcal{U}_{+}$ & The set of selected clients for federated training. \\
$\mathcal{L}_{u}$ & The local loss function of the user $u$. \\
$t$ & The federated training round. \\
$g(\cdot)$ & The client selection function. \\
$\mathcal{U}_{+}^{(t)*}$ & The optimal subset of clients at round $t$. \\
$\hat{\mathcal{L}}_{u}$ & the predicted triplet losses. \\
$r_{t}$ & The computed reward by the agent.  \\
\bottomrule[1.5pt] 
\end{tabular}
\label{tab:notations}
\end{table}

\subsection{Federated Recommender Systems}
Federated recommender systems (FedRSs) typically involve two parties: clients and a central server. Let $\mathcal{U}$ and $\mathcal{V}$ denote the set of users and items, respectively. On the client side, each user $u \in \mathcal{U}$ is associated with a unique client\footnote{Throughout the paper, we use the terms USER and CLIENT interchangeably depending on the context.}, which locally stores the user-item interaction dataset $\mathcal{D}_{u}=\{(u,v)\}_{v \in \mathcal{V}_{u}}$, where $\mathcal{V}_{u} \subseteq \mathcal{V}$ is the set of items that the user $u$ has interacted with. In addition, each client maintains a local model $f_{u}(\textbf{u},P_{u},\Theta_{u})$ parameterized by the user embedding $\textbf{u}$, local item embedding table $P_{u}$, and local model parameter $\Theta_{u}$\footnote{In recommender systems, mainstream models are typically embedding-based. Therefore, throughout this paper, when we refer to model parameters, we specifically mean non-embedding parameters, such as the weight parameters in deep neural networks.}. Importantly, in the FedRS setting, user private data, including the user-item interaction dataset $\mathcal{D}_{u}$ and the user embedding $u$, is strictly retained on local devices. It is neither uploaded to the server nor shared with other users, thereby ensuring strong data privacy.
On the server side, there is a global model $f_{s}(P_{s},\Theta_{s})$ parameterized by global item embedding table $P_{s}$ and the global model parameters $\Theta_{s}$. Typically, the training pipeline of a federated recommender system is as follows:
\begin{itemize}
    \item \textbf{Step 1 (Client Selection)}: In each federated training round $t$, based on a client selection strategy, the server selects a subset of clients $\mathcal{U}_{+}^{(t)} \subseteq \mathcal{U}$ to participate in federated training. The global item embedding table $P_{s}^{(t-1)}$ and global model parameters $\Theta_{s}^{(t-1)}$ from the previous round are then distributed to the selected clients, serving as initialization, for their local item embedding tables and model parameters in round $t$ of local training, as follows:
    \begin{equation}
    P_u^{(t)},\Theta_u^{(t)} \leftarrow  P_s^{(t-1)}, \Theta_s^{(t-1)}, \quad \forall u \in \mathcal{U}^{(t)}_{+}
    \end{equation}
    \item \textbf{Step 2 (Local Model Training)}: Each selected client performs local model training based on its local dataset, updating the user embeddings, local item embedding table, and local model parameters, as follows:
    \begin{equation}
    \textbf{u}^{(t)}, P_u^{(t)},\Theta_u^{(t)} \leftarrow  f_{u}(\mathcal{D}_{u}), \quad \forall u \in \mathcal{U}^{(t)}_{+}
    \end{equation}
    \item \textbf{Step 3 (Global Model Update)}: After local training, the updated local item embedding tables and local model parameters are sent back to the server. The server then aggregates the collected local item embedding tables and model parameters using a model aggregation algorithm, such as $FedAvg(\cdot)$ \cite{mcmahan2017communication}, to update the global item embedding table and global model parameters, as follows:
    \begin{equation}
    P_s^{(t)},\Theta_s^{(t)} = FedAvg\left(\left\{P_u^{(t)},\Theta_u^{(t)} \mid u \in \mathcal{U}^{t}_{+} \right\}\right)
    \end{equation}
\end{itemize}
Steps 1–3 are repeated iteratively until the termination condition is met, ensuring that the global loss function is minimized as the training progresses.
\begin{equation}
\min \sum_{u \in \mathcal{U}} \mathcal{L}_u
\end{equation}
where $\mathcal{L}_{u}$ is the local loss function of the user $u$.

\subsection{Problem Formulation: Client Selection in FedRSs}
In federated recommender systems (FedRSs), the client selection problem focuses on determining an optimal subset of clients to participate in each training round to improve the global model’s performance. 

\textbf{Input:} Candidate client set $\mathcal{U}$, where each client $u \in \mathcal{U}$ is associated with non-private information (e.g., local dataset size, or previous training loss,) that can be safely shared with the server.

\textbf{Output:} A selected subset of clients $\mathcal{U}_{+}^{(t)} \subseteq \mathcal{U}$ for round $t$, where the mapping is defined by a client selection function $g(\cdot)$ :
\begin{equation}
    \mathcal{U}_{+}^{(t)} = g(\mathcal{U}, \{\phi_u \mid u \in \mathcal{U}\}),
\end{equation}

\textbf{Objective:} he goal is to select an optimal subset $\mathcal{U}_{+}^{(t)*}$ that minimizes the expected global loss after aggregation:
\begin{equation}
    \mathcal{U}_{+}^{(t)*} :=
\underset{\mathcal{U}_{+}^{(t)} \subseteq \mathcal{U}}{\arg\min} 
\; \sum_{u \in \mathcal{U}} \mathcal{L}_u,
\end{equation}

\textbf{Problem Complexity:} The solution space of this problem is combinatorially large, containing $\binom{|\mathcal{U}|}{|\mathcal{U}_{+}^{(t)}|}$ possible client subsets in each training round. Exhaustive search is computationally infeasible for real-world FedRSs with millions of clients. Moreover, due to strict privacy constraints, user interaction data remains on local devices, making it impossible for the server to directly evaluate each client’s contribution to the global loss before selection. 

\begin{figure*}
    \centering
    \includegraphics[width=1\linewidth]{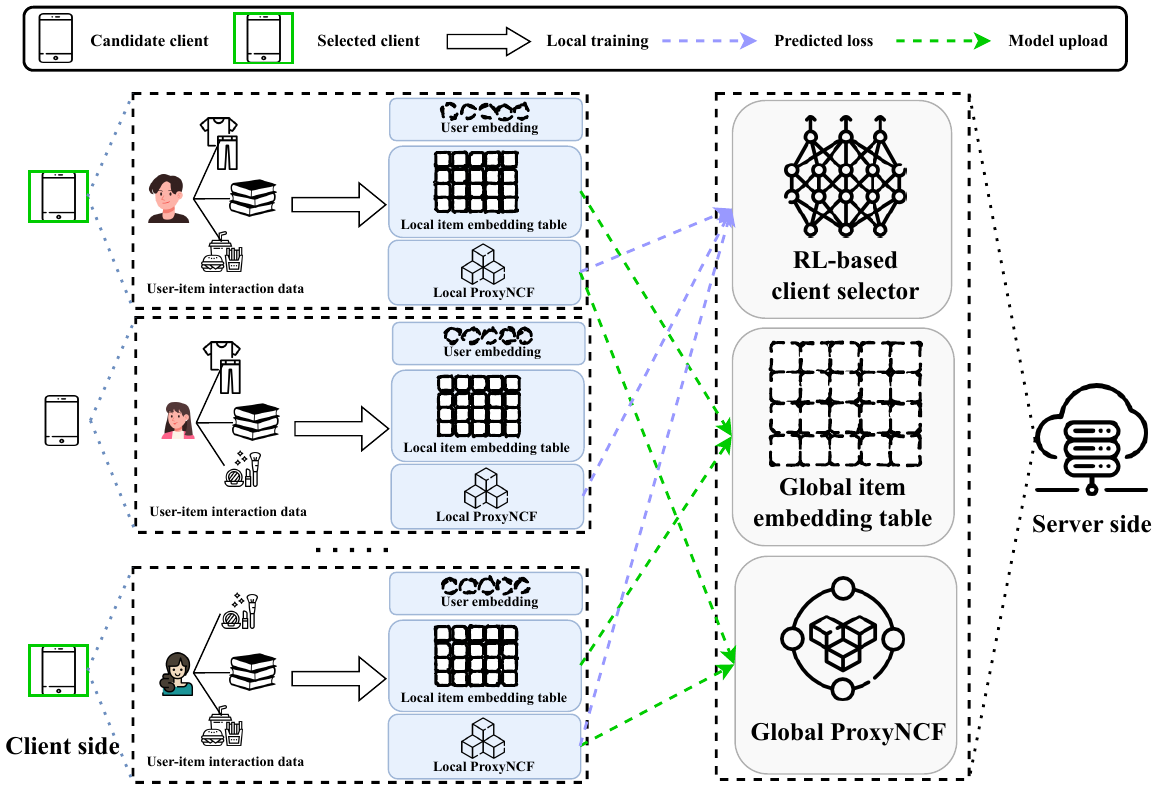}
    \caption{The architecture of the proposed ProxyRL-FRS}
    \label{fig:ProxyRL-FRS}
\end{figure*}

\section{ProxyRL-FRS}
In this section, we elaborate on our proposed framework, ProxyRL-FRS, as illustrated in Figure~\ref{fig:ProxyRL-FRS}. The framework consists of client-side and server-side modules. On the client side, each client strictly retains its private user--item interaction data locally, without sharing it with the server to ensure privacy. Each client also maintains a local model composed of three components: (i) a \textbf{user embedding}, which serves as a user representation encoding personalized preferences, and is strictly stored and updated locally for privacy protection; (ii) a \textbf{local item embedding table}, which stores item representations specific to each client. In each training round, selected clients download the global item embeddings from the server as initialization and update them locally using their private interaction data. After local training, the updated item embeddings are uploaded back to the server for aggregation; and (iii) a \textbf{local ProxyNCF model}, which consists of two branches: a NCF branch, responsible for standard recommendation model training, and a proxy branch for lightweight contribution prediction. At the beginning of each round, the proxy branch performs a single inference pass to produce predicted contributions (e.g., predicted loss), which are uploaded to the server to facilitate client selection. The detailed architecture of ProxyNCF are described in Section~\ref{sec:proxyncf}.

On the server side, it includes three key components: (i) an \textbf{RL-based client selector}, which selects participating clients in each round based on the predicted loss by their local ProxyNCF models. The detailed architecture of the client selector are detailed in Section~\ref{sec:rlselector}; (ii) a \textbf{global item embedding table}, which aggregates the uploaded item embeddings from selected clients and redistributes the updated embeddings in the next round; and (iii) a \textbf{global ProxyNCF model}, which aggregates model parameters from selected clients. It is worth noting that, to preserve user privacy, the server does not store any raw user data, and communication between the server and clients only involves exchanging model parameters. 

\subsection{Client Side}\label{sec:proxyncf}
Each client locally maintains three components: a user embedding, a local item embedding table, and the ProxyNCF model, which are described in detail in the following subsections.
\subsubsection{User Embedding}
Each client maintains a personalized user embedding vector $\mathbf{u} \in \mathbb{R}^d$ that represents its individual user preference. This embedding is generated locally by mapping the client’s unique ID into a $d$-dimensional latent space:
\begin{equation}
    \mathbf{u} = \mathrm{Embed}_U(\mathrm{ID}_u)
\end{equation}
where $\mathrm{Embed}_U(\cdot)$ is a local embedding lookup function. Importantly, this user embedding is strictly stored and updated on the client side and is never shared with the serve

\subsubsection{Local Item Embedding Table}
In addition to the user embedding, each client maintains a local item embedding table $P_u \in \mathbb{R}^{|\mathcal{V}| \times d}$.
At the beginning of each training round $t$, selected clients receive the global item embeddings $P_s^{(t-1)}$ from the server as initialization:
\begin{equation}
    P_u^{(t)} \leftarrow P_s^{(t-1)}.
\end{equation}
During local training, the item embeddings are updated based on the client’s private user-item interaction data:
\begin{equation}
    \mathbf{v} \leftarrow \mathbf{v} - \eta \frac{\partial \mathcal{L}_{NCF}}{\partial \mathbf{v}}, \quad \forall v \in \mathcal{V}_u,
\end{equation}
where $\mathcal{V}_u$ is the set of items that user $u$ has interacted with. 
After training, the updated local item embedding table $P_u^{(t)}$ is uploaded back to the server to facilitate subsequent global model updates.
It is worth noting that since each user only updates the embeddings of items they have interacted with, the server could potentially infer a user’s interaction history from the uploaded item embeddings. To mitigate this privacy risk, the proposed method can be seamlessly integrated with local differential privacy (LDP) \cite{truex2020ldp} by adding Gaussian noise to each item embedding table before uploading:
\begin{equation}
\tilde{P}_u = P_u + \mathcal{N}(0, \sigma^2 I),
\end{equation}
where $\sigma$ denotes the standard deviation controlling the noise level, and $I$ is the identity matrix.

\subsubsection{ProxyNCF Model}
\begin{figure}
    \centering
    \includegraphics[width=0.8\linewidth]{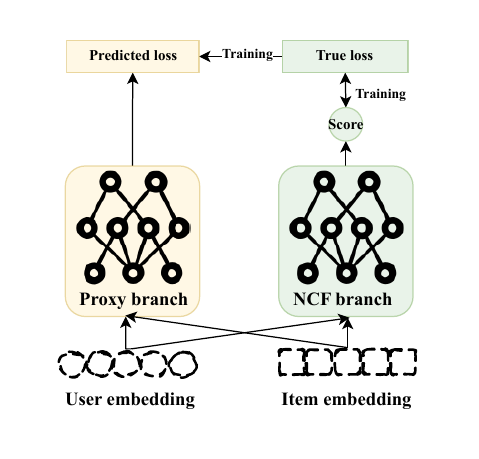}
    \caption{The architecture of the proposed ProxyNCF.}
    \label{fig:ProxyNCF}
\end{figure}

As previously discussed, a major technical challenge in client selection for federated recommender systems lies in the expensive contribution evaluation, resulting from the large number of participating clients. This makes directly applying existing generic client selection methods non-trivial, as they typically require pre-round local training on each client to obtain accurate contribution estimates, leading to significant computational overhead and substantial time consumption.
To address this challenge, we propose \textit{ProxyNCF}, a dual-branch neural model that extends the standard Neural Collaborative Filtering (NCF) architecture with an additional proxy branch. The proxy branch enables efficient prediction of each client’s contribution without requiring time-consuming pre-round local training. As shown in Figure~\ref{fig:ProxyNCF}, ProxyNCF is composed of two branches: (i) an NCF branch responsible for training the local recommendation model, and (ii) a proxy branch dedicated to predicting contribution scores (e.g., expected training loss), thereby facilitating efficient client selection on the server side.

\paragraph{NCF Branch}
Since the NCF branch is primarily responsible for training the standard recommendation model on the selected clients in each round, 
we adopt the original NCF architecture \cite{he2017neural} \footnote{As this work does not focus on developing new backbone recommendation models, 
we use NCF as a representative example due to its widespread adoption. 
Nevertheless, our framework is model-agnostic and can be seamlessly integrated with other backbone recommendation models.}. 
Specifically, it embeds user $u$ and item $v$ into $d$-dimensional vectors $\mathbf{u} \in \mathbb{R}^d$ and $\mathbf{v} \in \mathbb{R}^d$, respectively. 
The predicted similarity score between user $u$ and item $i$ is:
\begin{equation}
    \hat{y}_{u,i} = \mathrm{MLP}\left([\mathbf{u} \| \mathbf{v}]\right),
\end{equation}
where $[\cdot \| \cdot]$ denotes concatenation and $\mathrm{MLP}(\cdot)$ is a multi-layer perceptron (MLP) defined as:
\begin{equation}
    \mathrm{MLP}(\mathbf{x}) = W_2 \cdot \sigma(W_1 \mathbf{x} + \mathbf{b}_1) + \mathbf{b}_2,
\end{equation}
with $W_1 \in \mathbb{R}^{h \times 2d}$, $W_2 \in \mathbb{R}^{1 \times h}$, $\mathbf{b}_1 \in \mathbb{R}^{h}$, $\mathbf{b}_2 \in \mathbb{R}$, 
and $\sigma(\cdot)$ the ReLU activation, where $h$ denotes the dimension of the hidden layer. 
We adopt the widely used Bayesian Personalized Ranking (BPR) loss \cite{rendle2012bpr} to optimize the local NCF branch for each user \( u \), which aims to maximize the ranking score difference between positive and negative items, thereby improving the personalized ranking quality:
\begin{equation}
\label{equ:updatencfbranch}
\begin{aligned}
    \mathcal{L}_{NCF} &= \sum_{(u,v,j) \in \mathcal{D}_{u}} \ell_{u,v,j}, \\
    \text{where} \quad
    \ell_{u,v,j} &= - \ln \sigma\left( \hat{y}_{u,v} - \hat{y}_{u,j} \right),
\end{aligned}
\end{equation}
where $v \in \mathcal{V}_u$ is a positive item that user $u$ has interacted with, $j \notin \mathcal{V}_u$  is a randomly sampled negative item, 
and $\ell_{u,v,j}$ represents the loss for a sampled triplet $(u, v, j)$ from the local dataset $\mathcal{D}_{u}$.

\paragraph{Proxy Branch}
To eliminate the need for time-consuming pre-round local training for contribution evaluation, 
we introduce a proxy branch to predict each client's training loss. 
The proxy branch shares the same input as the NCF branch, including the user embedding $\mathbf{u}$ and item embeddings $\mathbf{v}$, 
as well as the same MLP architecture. 
Despite sharing the same structure, the two branches are optimized independently with different objectives: 
the NCF branch is supervised by implicit feedback through pairwise ranking loss to improve recommendation accuracy, 
while the proxy branch is trained with ground-truth losses from the NCF branch to accurately estimate each client’s training loss. 
This decoupled optimization ensures that the proxy branch focuses solely on contribution estimation without interfering with the primary recommendation task.

Specifically, for each training triplet $(u, v, j)$, the proxy branch outputs a predicted pairwise loss $\hat{\ell}_{u,v,j}$ 
as an estimation of the true training loss $\ell_{u,v,j}$:
\begin{equation}
    \hat{\mathcal{L}}_{u} = \sum_{(u,v,j) \in \mathcal{D}_{u}} \hat{\ell}_{u,v,j},
    \label{equ:proxyloss}
\end{equation}
and is trained using a regression loss that minimizes the mean squared error (MSE) between the predicted and actual loss values:
\begin{equation}
\label{equ:updateproxybranch}
    \mathcal{L}_{\text{Proxy}} = \sum_{(u,v,j) \in \mathcal{D}_{u}} \left(\hat{\ell}_{u,v,j} - \ell_{u,v,j} \right)^2.
\end{equation}

\begin{figure*}
    \centering
    \includegraphics[width=0.95\linewidth]{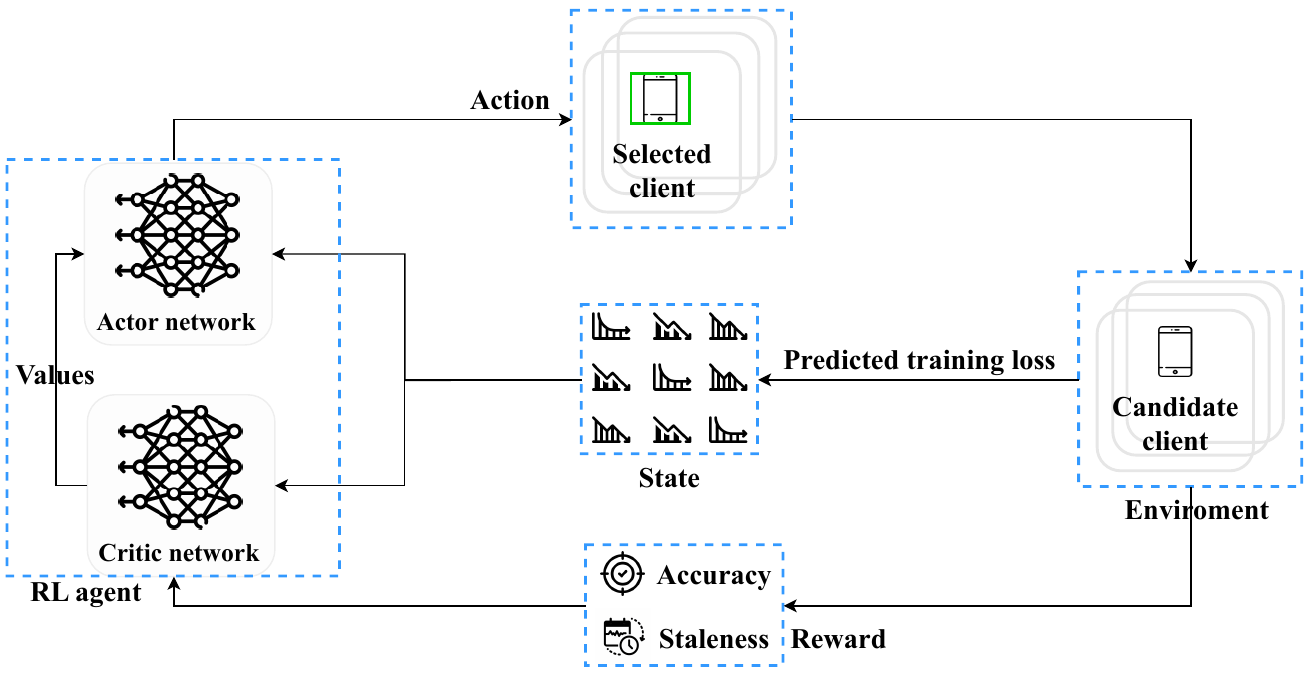}
    \caption{The architecture of the proposed staleness-aware RL agent.}
    \label{fig:rlagent}
\end{figure*}

\subsection{Server Side}

On the server side, ProxyRL-FRS comprises three key components: (1) a staleness-aware reinforcement learning (RL) agent for client selection, (2) a global item embedding table for maintaining collaborative item representations; (3) a global ProxyNCF model for aggregating model parameters.

\subsubsection{Staleness-Aware RL Agent for Client Selection}\label{sec:rlselector}

A unique technical challenge in federated recommender systems is the issue of \textit{sparse updates}, which fundamentally differs from generic federated learning tasks. In generic federated learning applications, model parameters are primarily composed of dense neural network weights, and local training on each client updates nearly all of these parameters.
In contrast, recommendation models are typically dominated by a large item embedding table that stores item representations. Due to the long-tail nature of user–item interactions and strict privacy constraints that prevent sharing interaction data with the server, each client only updates the embeddings corresponding to its own local interaction history. The server, therefore, cannot directly identify which clients are associated with stale item embeddings. As a result, even when many clients participate in federated training, only a small fraction of embeddings are updated in each round.
This leads to highly \textit{sparse updates}, causing many item embeddings—especially those in the long tail—to remain stale for extended periods. Such embedding staleness degrades recommendation quality and slows global model convergence.

To address these challenges, we propose a staleness-aware RL agent built upon the actor–critic framework. Unlike existing RL-based approaches that focus solely on recommendation accuracy as the reward signal, our RL agent introduces a novel reward function that explicitly penalizes embedding staleness. By incorporating a staleness-aware term into the reward, the agent adaptively favors clients that can refresh long-tail or rarely updated item embeddings, thereby mitigating the sparse update problem. The entire selection process operates solely on non-private signals and aggregated staleness statistics, ensuring strict privacy preservation. The key components are illustrated in Figure \ref{fig:rlagent} and detailed below.
\begin{itemize}
    \item \textbf{State ($s^{(t)}$)}: 
    We represent the state in each communication round by evaluating the contribution of every client. 
    Since client-side training loss has been widely used as an indicator of client importance in federated learning~\cite{cho2022towards,li2024comprehensive}, 
    we use the predicted training loss of each client by the proxy branch as a practical signal. 
    The motivation is twofold. First, the local training loss provides a straightforward measure of how well each client's data is fitted by the current global model: a larger loss suggests the client’s update could be more beneficial for model improvement. Second, using proxy or predicted losses (rather than true losses computed on all client data) significantly reduces the computation and communication required at each round, making the proposed method more scalable in practice. Formally, at round $t$, the state is defined as
    \begin{equation}
        s^{(t)} = \{\hat{\mathcal{L}}_u\}_{u \in \mathcal{U}},
        \label{equ:state}
    \end{equation}
    where $\hat{\mathcal{L}}_u$ denotes the proxy-estimated training loss of client $u$.

    \item \textbf{Action ($a^{(t)}$)}: The action is to select a subset of clients $\mathcal{U}_{+}^{(t)} \subseteq \mathcal{U}$ to participate in local training. 
    
    \item \textbf{Reward ($r^{(t)}$)}: To jointly optimize recommendation performance and mitigate sparse updates caused by embedding staleness, 
    we design the reward function as:
    \begin{equation}
        r^{(t)} = \lambda \cdot \text{Acc}^{(t)} - (1 - \lambda) \cdot \text{Staleness}^{(t)},
        \label{equ:reward}
    \end{equation}
    where $\text{Acc}^{(t)}$ is the recommendation accuracy (e.g., HR@10 or NDCG@10) on the validation set, 
    $\text{Staleness}^{(t)}$ quantifies the average staleness of item embeddings at round $t$, 
    and $\lambda \in [0,1]$ is a trade-off parameter that balances accuracy improvement and embedding staleness:
    \begin{equation}
        \text{Staleness}^{(t)} = \frac{1}{|\mathcal{V}|} \sum_{v \in \mathcal{V}} \frac{\tau^{(t)}(v)}{T}.
        \label{equ:staleness}
    \end{equation}
    Here, $\tau^{(t)}(v)$ denotes the number of rounds since item $v$ was last updated, and $T$ is a normalization constant (e.g., a staleness window). 
    By penalizing high staleness values, the RL agent is encouraged to select clients that can refresh long-tail or rarely updated item embeddings, thus alleviating the sparse update problem and accelerating model convergence.

    \item \textbf{Environment}: The environment corresponds to the federated recommender system, which updates the global model based on the selected clients’ contributions and provides the next state and reward signal after each communication round.    
    \item \textbf{Actor Network.}  
    The actor network takes the current state $s^{(t)}$, representing the proxy-predicted losses of all clients, as input. 
    It uses a multi-layer perceptron with ReLU activations followed by a softmax layer to produce a probability distribution $\pi(a^{(t)} \mid s^{(t)})$ 
    over candidate clients. A subset $\mathcal{U}_+^{(t)}$ is then sampled from this distribution for local training.
    
    \item \textbf{Critic Network.}  
    The critic network takes the same state $s^{(t)}$ as input and outputs a scalar value $V(s^{(t)})$ representing the expected cumulative reward. 
    It is implemented as an MLP with ReLU activations and a linear output layer.Both actor and critic networks are jointly updated using policy gradient and temporal-difference learning 
    \cite{fujimoto2018addressing    }, driven by the staleness-aware reward $r^{(t)}$.
\end{itemize}

\subsubsection{Global Model Update}
Once the selected clients $\mathcal{U}^{(t)}_{+}$ finish local training, 
they return the updated gradients of their ProxyNCF models and local item embeddings to the server. 
The server aggregates these updates using the standard FedAvg algorithm \cite{mcmahan2017communication} 
to update both the global ProxyNCF parameters and the global item embedding table:
\begin{equation}
\label{equ:globalupdate}
\begin{aligned}
    P_s^{(t)} &= P_s^{(t-1)} - \eta \cdot 
    \frac{1}{\sum_{u \in \mathcal{U}^{(t)}_{+}} |\mathcal{D}_u|} 
    \sum_{u \in \mathcal{U}^{(t)}_{+}} |\mathcal{D}_u| \cdot \nabla P_u^{(t)}, \\
    \Theta_s^{(t)} &= \Theta_s^{(t-1)} - \eta \cdot 
    \frac{1}{\sum_{u \in \mathcal{U}^{(t)}_{+}} |\mathcal{D}_u|} 
    \sum_{u \in \mathcal{U}^{(t)}_{+}} |\mathcal{D}_u| \cdot \nabla \Theta_u^{(t)},
\end{aligned}
\end{equation}
where $\eta$ is the learning rate for global model updates.

\begin{algorithm}[t]
\caption{ProxyRL-FRS}
\label{alg:proxylfrs}
\begin{algorithmic}[1]
\REQUIRE Local datasets $\{\mathcal{D}_u\}_{u \in \mathcal{U}}$, initial global ProxyNCF model $\Theta_s^{(0)}$, item embedding table $P_s^{(0)}$, RL actor $\pi_\theta$, critic $V_\psi$, total rounds $T$

\FOR{each round $t = 1$ to $T$}
\STATE {\color{blue} \textbf{Client Side: Predict client contributions}}
    \FOR{each client $u \in \mathcal{U}$}
        \STATE Compute predicted loss $\hat{\mathcal{L}}_u$ \COMMENT{cf. Eq.~\ref{equ:proxyloss}}
    \ENDFOR
\STATE {\color{blue} \textbf{Server Side: Client selection}}
    \STATE Server constructs state $\mathbf{s}_t = \{\hat{\mathcal{L}}_u\}_{u \in \mathcal{U}}$
    \STATE Actor samples client subset $\mathcal{U}_+^{(t)} \sim \pi_\theta(\cdot | \mathbf{s}_t)$

\STATE {\color{blue} \textbf{Client Side: Local training}}
    \FOR{each selected client $u \in \mathcal{U}_+^{(t)}$}
        \STATE Update NCF branch  \hfill \COMMENT{cf. Eq.~\ref{equ:updatencfbranch}}
        \STATE Update proxy branch \hfill \COMMENT{cf. Eq.~\ref{equ:updateproxybranch}}
    \ENDFOR

\STATE {\color{blue} \textbf{Server Side: Global model updates}}
    \STATE Server aggregates updates \hfill \COMMENT{cf. Eq.~\ref{equ:globalupdate}}
    \STATE Server updates RL agent based on the computed reward
\ENDFOR
\end{algorithmic}
\end{algorithm}

\subsection{Training Process}
Algorithm~\ref{alg:proxylfrs} presents the overall training procedure of ProxyRL-FRS. In each communication round, the training starts with each client predicting its own training loss using the proxy branch (line 3-5). These proxy-estimated losses are collected and used by the server to construct the state representation for the staleness-aware RL agent (line 7). Based on this state, the actor network samples a subset of clients to participate in the current round of federated training (line 8).
Each selected client then performs local training on its private data. Specifically, it updates the NCF branch using BPR loss and simultaneously supervises the proxy branch using ground-truth losses (line 10–13). After local updates, clients send their updated model parameters and item embeddings to the server.
The server aggregates these updates using FedAvg to update both the global ProxyNCF and the global item embedding table (line 15). Lastly, the server updates the RL agent based on the computed reward (line 16). This process is repeated until the convergence criterion is met.

\section{Experiments}
To validate the effectiveness of our proposed method, we conducted experiments on three public datasets, aiming to answer the following research questions:
\begin{itemize}
    \item \textbf{RQ1:} How does the proposed method compare to state-of-the-art federated recommendation methods in terms of recommendation quality?
    \item \textbf{RQ2:} Does the proposed method train faster and more efficiently than state-of-the-art federated recommendation methods?
    \item \textbf{RQ3:} How do the key components of the proposed model contribute to its overall performance?
    \item \textbf{RQ4:} How do different client selection methods compare in terms of selection bias, such as user coverage and participation frequency?
    \item \textbf{RQ5:} How do different hyperparameter settings affect the performance of the proposed model?
\end{itemize}
\vspace{-1.5em}
\subsection{Datasets} 
We validated the performance of our model on three widely-used public datasets, covering different domains and varying levels of sparsity, as detailed below:
\begin{itemize}
    \item \textbf{MovieLens-1M}\footnote{https://grouplens.org/datasets/movielens/} is the dataset containing user ratings for movies. Each user has rated at least 20 movies. 
    \item \textbf{Fashion} \cite{ni2019justifying} is the dataset containing user reviews for fashion products on Amazon. We filter out users with fewer than 20 ratings.
    \item \textbf{Video Games \cite{ni2019justifying}} is the dataset containing user reviews for video games products on Amazon. We filter out users with fewer than 10 ratings.
\end{itemize}
Each dataset is randomly split into training, validation, and test sets in a ratio of 8:1:1.
The statistical information of these datasets is summarized in Table \ref{tab:datasets}: 

\begin{table}[!htbp]
  \centering
  \caption{Statistics of datasets}
    \begin{tabular}{c|c|c|c|c}
    \toprule
    Dataset & \# Users & \# Items & \# Interactions & Sparsity \\
    \midrule
    \midrule
    MovieLens-1M & 6,040 & 3,900 & 1,000,209 & 95.75\% \\
    Fashion & 29,858 & 40,981 & 1,027,370 & 99.92\% \\
    Video Games & 15,517 & 37,077 & 284,867 & 99.95\% \\
    \bottomrule
    \end{tabular}
\label{tab:datasets}
\end{table}

\begin{table*}[htbp]
  \centering
  \caption{Performance comparison of various client selection methods on three datasets.
The best results are highlighted in \textbf{bold}, while the best-performing baselines are \underline{underlined}. All improvements are statistically significant with $p < 0.05$.}
    \begin{tabular}{c|l|cc|cc|cc}
    \toprule
    \multicolumn{2}{c|}{Dataset} & \multicolumn{2}{c|}{MovieLens-1M} & \multicolumn{2}{c|}{Fashion} & \multicolumn{2}{c}{Video Games} \\
    \midrule
    \multicolumn{2}{c|}{Evaluation metrics} & HR@20 & NDCG@20 & HR@20 & NDCG@20 & HR@20 & NDCG@20 \\
    \midrule
    \midrule
    \multicolumn{1}{c|}{\multirow{5}[2]{*}{\makecell{Random \\CS}}} & FedMF & 0.4363 & 0.0652 & 0.0416 & 0.0056 & 0.2315 & 0.0163 \\
          & FedNCF & 0.4394 & \underline{0.0654} & 0.0447 & 0.0057 & 0.237 & 0.016 \\
          & PFedRec  & 0.4409 & 0.0653 & 0.0455 & 0.0059 & 0.2312 & 0.0162 \\
          & PFedGNN & 0.445 & 0.0596 & 0.0436 & 0.0056 & 0.229 & 0.0159 \\
          & GPFedRec & 0.4382 & 0.0652 & 0.0427 & 0.0055 & 0.2425 & 0.0176 \\
    \midrule
    \multicolumn{1}{c|}{\multirow{6}[2]{*}{\makecell{Non-random \\CS}}} & FedFast & 0.4399 & 0.0655 & 0.0405 & 0.0058 & 0.2467 & 0.0178 \\
          & Pow-d & \underline{0.4473} & 0.0643 & \underline{0.0494} & 0.0066 & \underline{0.2533} & \underline{0.0188} \\
          & FedRL & 0.4468 & 0.0653 & 0.0466 & 0.0066 & 0.2486 & 0.0181 \\
          & FedACS & 0.4462 & 0.0652 & 0.048 & \underline{0.0071} & 0.2503 & 0.0184 \\
          & FedGST & 0.4472 & 0.0638 & 0.048 & 0.0068 & 0.2494 & 0.0181 \\
          & \textbf{ProxyRL-FRS} & \textbf{0.4579} & \textbf{0.067} & \textbf{0.0576} & \textbf{0.0083} & \textbf{0.2596} & \textbf{0.0195} \\
    \bottomrule
    \end{tabular}%
  \label{tab:top-kresutls}%
\end{table*}%

\begin{figure*}[htbp]
    \centering
    \includegraphics[width=0.9\textwidth]{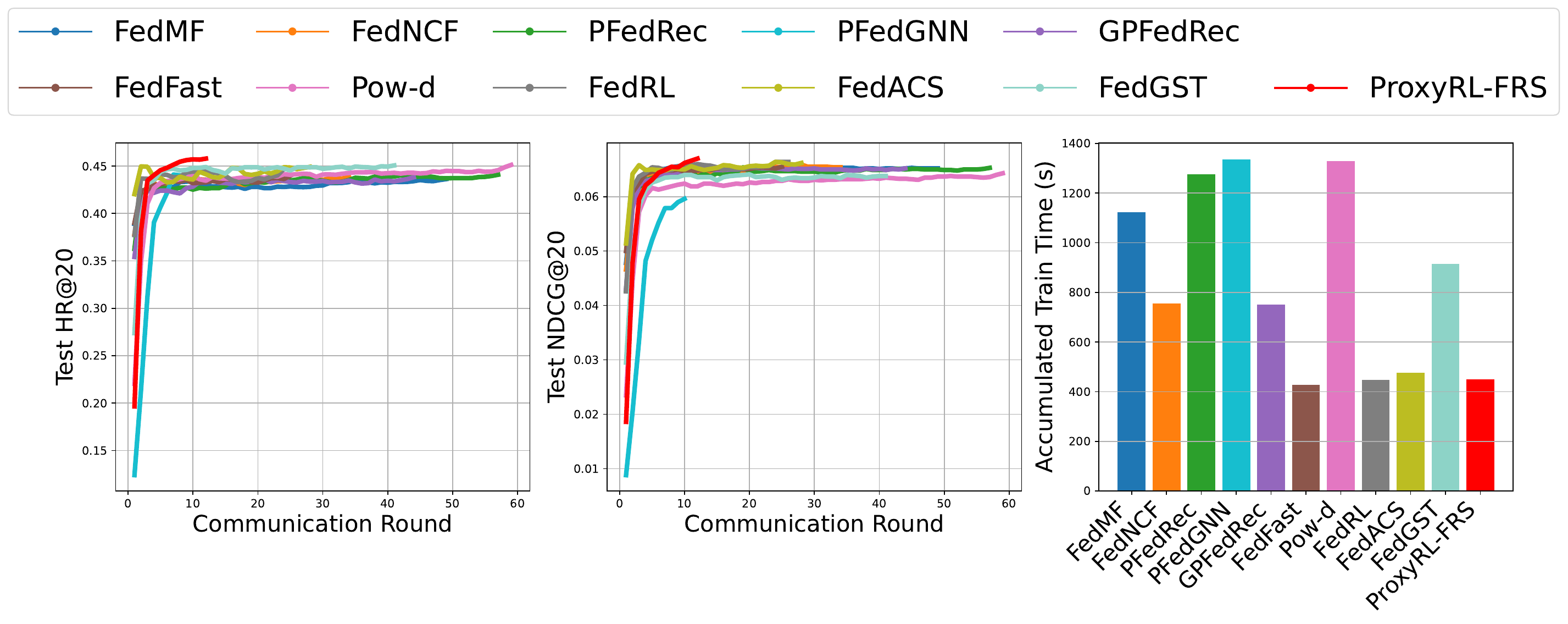}\\
    \includegraphics[width=0.85\textwidth]{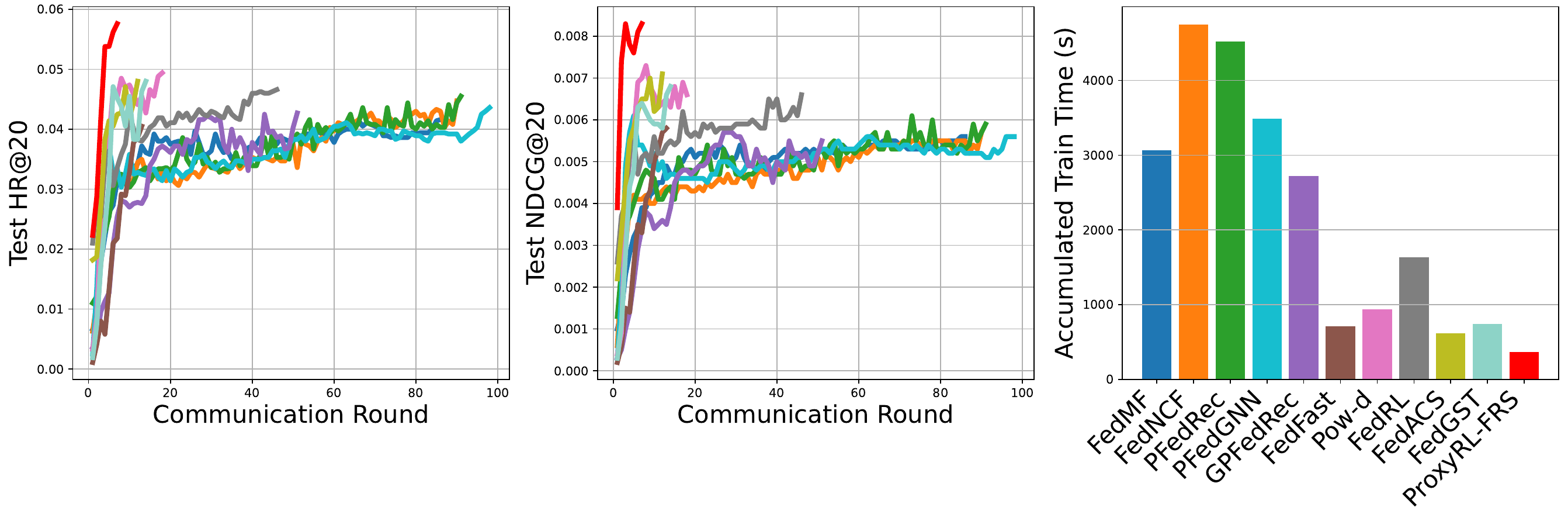}\\
    \includegraphics[width=0.85\textwidth]{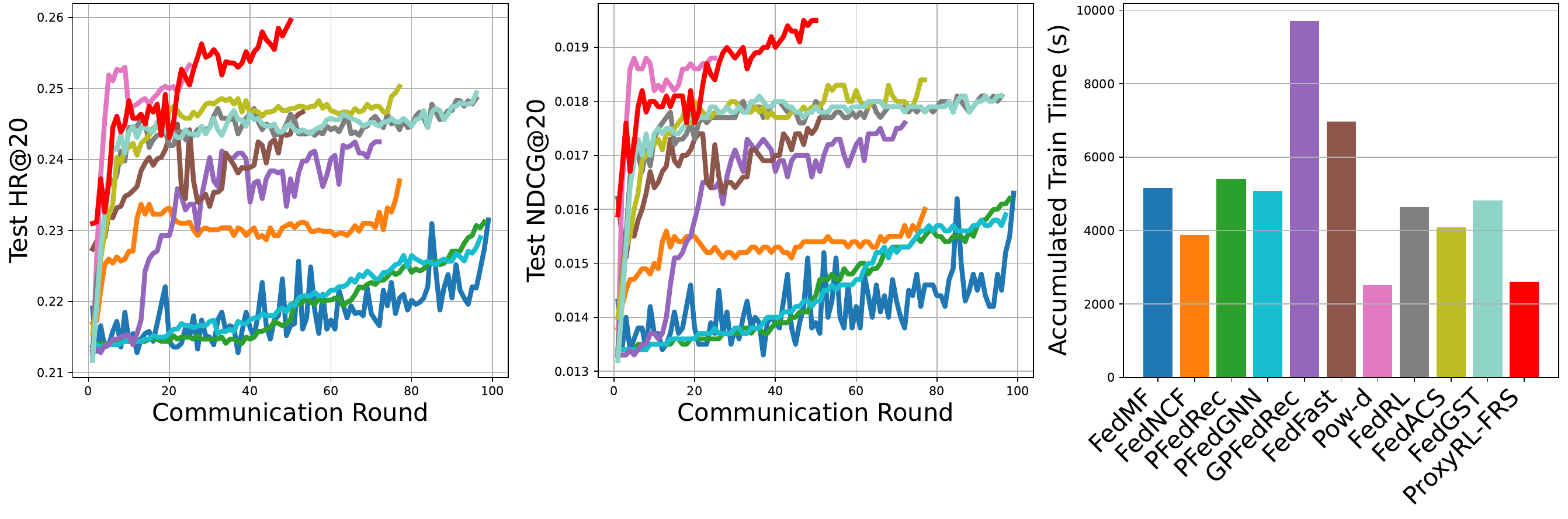}
    \caption{Comparison of model convergence and training efficiency across three datasets. Rows correspond to (top) MovieLens-1M, (middle) Fashion, and (bottom) Video Games. }
    \label{fig:efficiency}
\end{figure*}

\vspace{-2em}
\subsection{Baselines}
We compared our method with two types of client selection methods in federated recommender systems: (1) random client selection methods (Random CS), which select clients in a uniformly random manner in each round, and (2) non-random client selection methods (Non-random CS), which employ various mechanisms to guide the client selection process.
\begin{itemize}
    \item \textbf{Random CS methods:}
    \begin{itemize}
        \item \textbf{FedMF} \cite{chai2020secure} is a pioneering FedRS method that integrates matrix factorization into the federated learning framework to protect user data privacy.
        \item \textbf{FedNCF} \cite{perifanis2022federated} is a representative FedRS method that extends Neural Collaborative Filtering (NCF \cite{he2017neural}) to the federated setting. It introduces deep neural network layers to better capture complex user-item interaction patterns under privacy constraints.
        \item \textbf{PFedRec} \cite{10.24963/ijcai.2023/507} is a personalized FedRS method that learns user-specific score functions and item embedding table locally on each user's devices.
        \item \textbf{FedPerGNN} \cite{wu2022federated} is a graph neural network-based FedRS method that models user-item interaction data as ego-graphs on each user's device, while leveraging GNNs to capture collaborative signals over user-item bipartite graph.
        \item \textbf{GPFedRec} \cite{10.1145/3637528.3671702} is a graph-guided personalized FedRS method that constructs a user-relation graph based on the personalized item embedding tables.
    \end{itemize}
    \item \textbf{Non-random CS methods:}
    \begin{itemize}
    \item \textbf{FedFast} \cite{muhammad2020fedfast}: is a clustering-based client selection method that applies K-means to group users based on their meta-information, and then randomly selects clients within each cluster.
    \item \textbf{Pow-d} \cite{cho2022towards}: is a heuristic client selection method that introduces a Power-of-Choice strategy. It firstly allows candidate clients perform pre-round local training, and then selects those with higher local losses to participate in the next training round.
    \item \textbf{FedRL} \cite{di2024fedrl} is a reinforcement learning-based client selection method for federated recommendation. It employs a reinforcement learning agent to select clients based on indicators such as training loss, and utilizes a hypernetwork to generate personalized model parameters for local devices.
    \item \textbf{FedACS} \cite{xu2025fedacs} is a multi-armed bandit-based client selection method designed for federated recommendation. It leverages a global GNN model to learn embeddings of hidden edges among selected clients.
    \item \textbf{FedGST} \cite{tang2024fedgst} employs an influence function to evaluate contributions of candidate clients, and clients with high influence scores are selected to participate in the next federated training round.
    \end{itemize}
\end{itemize}

\vspace{-1em}
\subsection{Evaluation Metrics}
To evaluate recommendation accuracy, we adopt two widely used ranking metrics \cite{zhu2022bars}: HR@20 (Hit Ratio), which measures the proportion of positive items in the test set that appear in the top-20 predictions; and NDCG@20 (Normalized Discounted Cumulative Gain), which considers the positions of the positive items and assigns higher scores when they are ranked higher in the top-20 list.

\vspace{-1em}
\subsection{Experimental Setup}
For all model parameters, including user and item embeddings as well as the parameters of the deep neural networks, we use the Xavier method \cite{glorot2010understanding} for initialization. The embedding dimensions for users and items are set to 32 across all methods.
For the actor-critic networks, we adopt fully connected deep neural networks for both the actor and critic architectures. The activation function in the hidden layers of all models is set to ReLU($\cdot$), and we use Adam as the optimizer. For all models, we consider the training converged when either HR@20 on the validation set fails to improve for five consecutive rounds or the maximum number of training rounds (100 in our setting) is reached.

\subsection{Model Effectiveness (RQ1)}
We first compare our method with other baselines on the most common task in recommender systems, the Top-K recommendation task, where the goal is to recommend 
K items that are most likely to be of interest to the user. 
We report the performance of the models in terms of recommendation accuracy in Table \ref{tab:top-kresutls} 
From the experimental results, we observe the following:
\begin{itemize}
    \item Our method consistently achieves the highest recommendation accuracy across all three datasets, outperforming the state-of-the-art baselines. For example, on the extremely sparse Fashion dataset, our approach surpasses the best-performing baseline by 16.5\% and 16.7\% in HR@20 and NDCG@20, respectively, demonstrating its effectiveness.
    \item In most cases, non-random client selection methods yield higher recommendation accuracy compared to random selection, further confirming the necessity of effective client selection strategies.
    \item Our method delivers more significant improvements on sparser datasets (e.g., Fashion and Video Games) compared to other baselines. This is particularly important in recommender systems, where data sparsity is a common and challenging issue.
\end{itemize}

\subsection{Model Efficiency (RQ2)}
We compare the model convergence and training efficiency of all methods in Figure \ref{fig:efficiency}. From the results, we observe the following:
\begin{itemize}
    \item In most cases, our model achieves faster convergence while attaining the highest recommendation accuracy on the test set, demonstrating its ability to effectively balance accuracy and training efficiency.
    \item Although our method introduces additional training components, it maintains competitive overall efficiency. For example, on the Video Games dataset, our method’s training time is close to that of the fastest method, Pow-d, while achieving 2.5\% and 3.7\% higher performance in HR@20 and NDCG@20, respectively. This is reasonable because better client selection enables the model to converge faster, offsetting the extra computational overhead introduced by the auxiliary modules.
    \item In most cases, non-random client selection methods converge faster than random selection due to their biased client selection strategies. However, they also incur higher training costs, further validating the rationality and effectiveness of our proxy model-based approach.
    \end{itemize}

\begin{table*}[htbp]
  \centering
  \caption{Ablation Studies of ProxyRL-FRS.}
    \begin{tabular}{l|ccc|ccc|ccc}
    \toprule
    \multicolumn{1}{c|}{\multirow{2}[4]{*}{Method}} & \multicolumn{3}{c|}{MovieLens-1M} & \multicolumn{3}{c|}{Fashion} & \multicolumn{3}{c}{ Video Games} \\
\cmidrule{2-10}          & HR@20 & NDCG@20 & Train Time (s) & HR@20 & NDCG@20 & Train Time (s) & HR@20 & NDCG@20 & Train Time (s) \\
    \midrule
    \midrule
    w/o Proxy & 0.4572 & 0.0665 & 976.4 & 0.056 & 0.0077 & 1536.33 & 0.2584 & 0.0187 & 5396.4 \\
    w/o Staleness & 0.4512 & 0.0636 & 672.36 & 0.0459 & 0.0061 & 579.05 & 0.2432 & 0.0173 & 4125.33 \\
    w/o Accuracy  & 0.4387 & 0.0632 & 1324.2 & 0.0451 & 0.006 & 842.16 & 0.2379 & 0.0162 & 2871.05 \\
    \textbf{ProxyRL-FRS} & \textbf{0.4579} & \textbf{0.067} & \textbf{426.18} & \textbf{0.0576} & \textbf{0.0083} & \textbf{333.62} & \textbf{0.2596} & \textbf{0.0195} & \textbf{2375.16} \\
    \bottomrule
    \end{tabular}%
  \label{tab:Ablation study}%
\end{table*}%

\subsection{Ablation studies (RQ3)}
To assess the contribution of individual components to the model's performance, we conducted ablation experiments by analyzing how the removal of specific components affects the results. Specifically, we first evaluated the effect of removing the proxy model branch from ProxyNCF, which degrades the local model into a standard NCF model. In this case, each round requires standard pre-round training to obtain the local loss for client contribution evaluation. This variation is referred to as \textit{w/o Proxy.} Furthermore, we evaluated the performance when the staleness term was removed from the reward of the RL-based client selector, referred to as \textit{w/o Staleness.} Similarly, we tested the effect of removing the accuracy term from the reward, termed \textit{w/o Accuracy.}
The outcomes of these experiments are presented in Table \ref{tab:Ablation study}. Based on the results, we can draw the following observations:
\begin{itemize}
    \item Removing the proxy model branch from the local client resulted in nearly unchanged accuracy across the three datasets. However, the training time increased significantly, e.g., by 4.6× on the Fashion dataset. This is reasonable because directly computing the true client contribution requires extensive local training, which is time-consuming. On the other hand, despite using the true contribution values, the recommendation accuracy did not improve significantly. A plausible explanation is that the proxy model already provides sufficiently accurate approximations of client contributions, making the benefit of using the true values marginal compared to the additional computational cost.
    \item Removing the staleness term caused a notable performance drop: HR@20 decreased by 25.4\% and 6.7\% on the Fashion and Video Games datasets, respectively, while the MovieLens-1M dataset only experienced a 1.48\% drop. This is reasonable because Fashion and Video Games are extremely sparse datasets. Without the guidance of the staleness reward, long-tail items are rarely updated, leading to significant degradation in recommendation quality.
    \item Removing the accuracy term caused the most substantial accuracy drop across all three datasets. This is expected because the accuracy reward directly incentivizes the selection of clients with higher potential to improve the global model. Without this term, the RL-based selector cannot effectively distinguish high-contributing clients, leading to degraded overall performance.
\end{itemize}
\begin{figure*}[htbp]
    \centering
    \begin{minipage}[t]{0.35\textwidth}
        \centering
        \includegraphics[width=\linewidth]{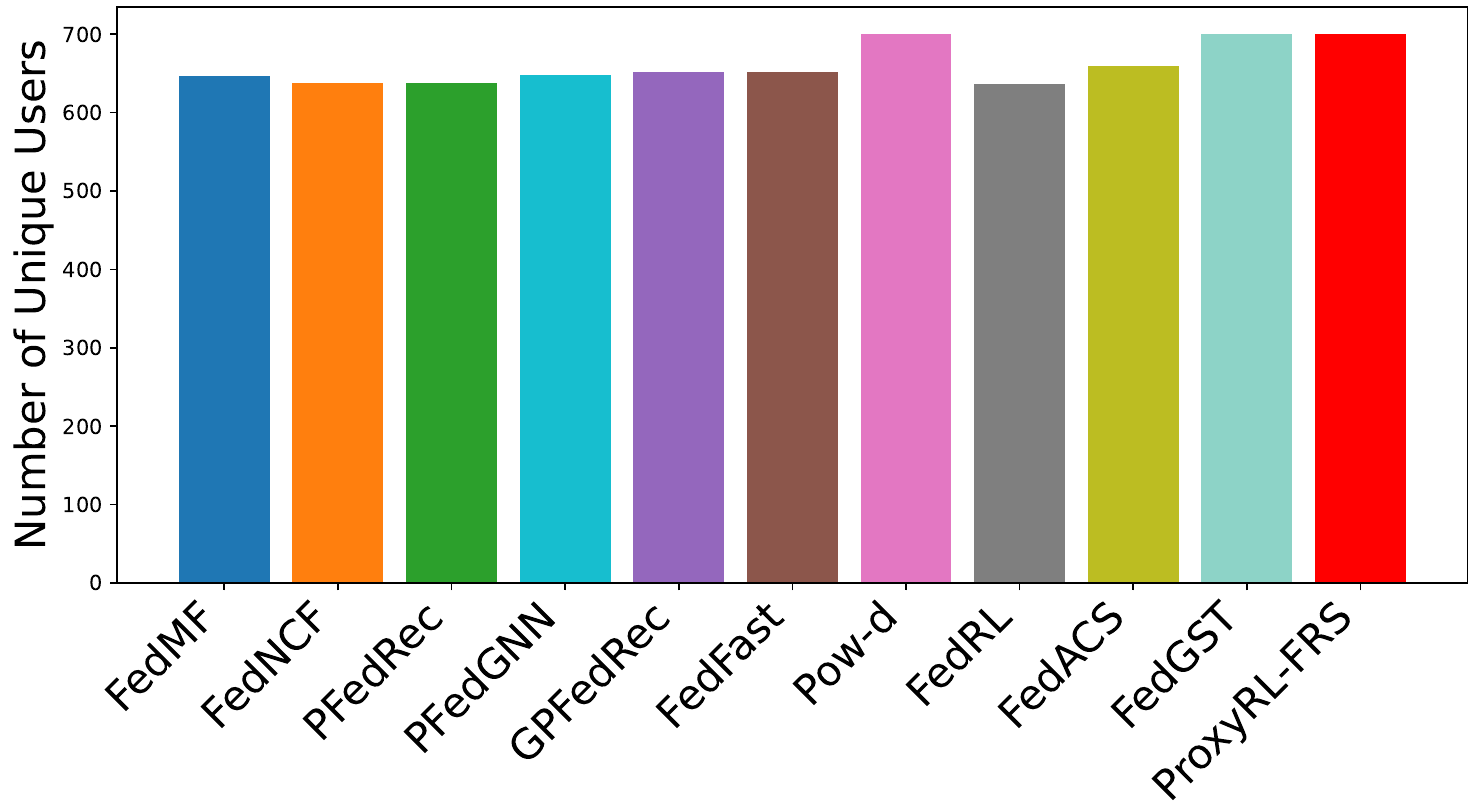}
        \caption{Number of unique selected clients.}
        \label{fig:uniqueclientnumber}
    \end{minipage}%
    \hfill
    \begin{minipage}[t]{0.65\textwidth}
        \centering
        \includegraphics[width=\linewidth]{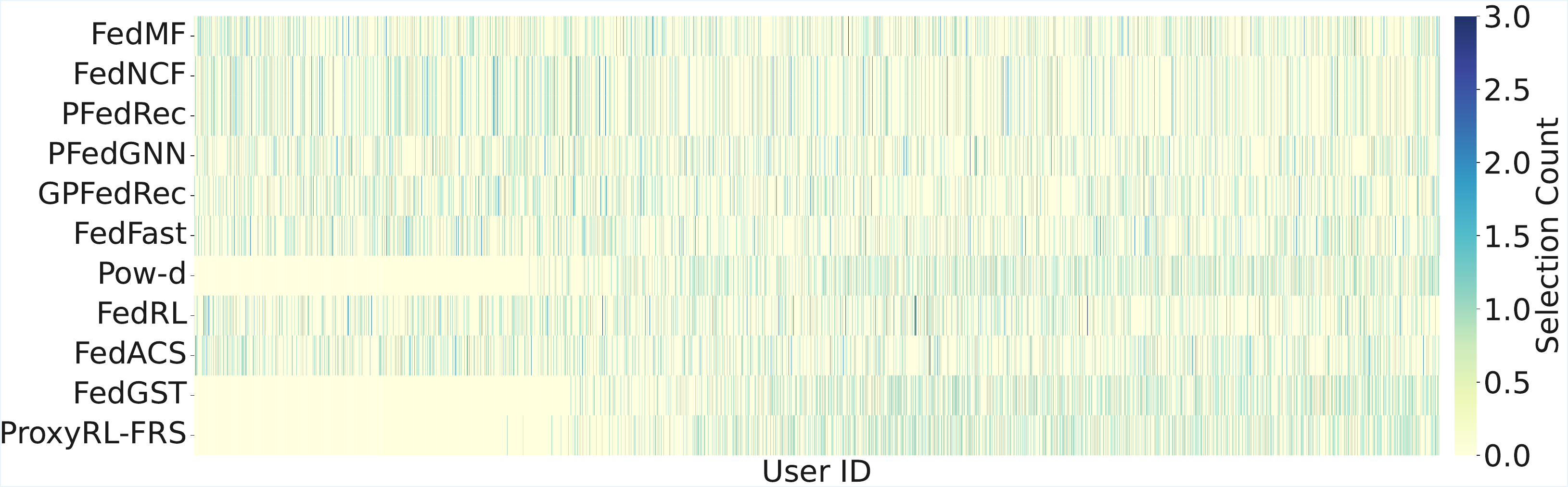}
        \caption{Client selection heatmap.}
        \label{fig:userheatmap}
    \end{minipage}
\end{figure*}

\begin{figure}
    \centering
    \includegraphics[width=1\linewidth]{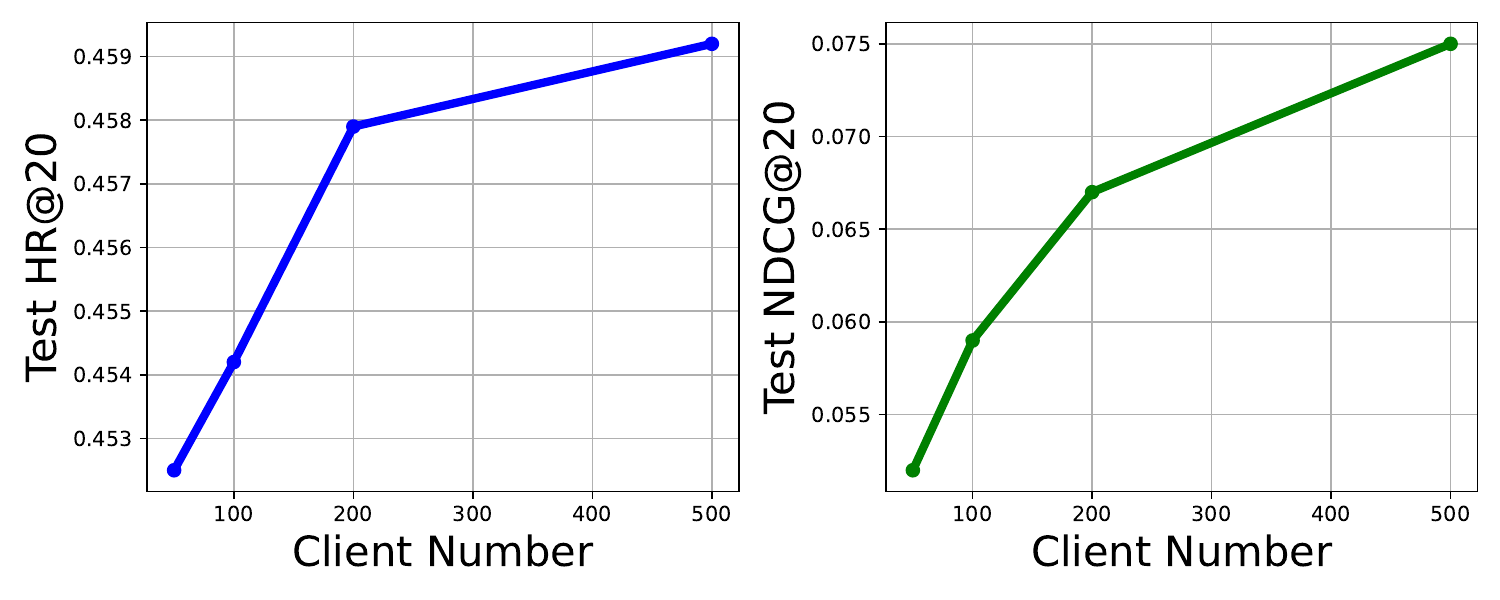}
    \caption{Model performance under different the number of selected clients $|\mathcal{U}_{+}|$ on MovieLens-1M dataset.}
    \label{fig:hyperparameter1}
\end{figure}
\begin{figure}
    \centering
    \includegraphics[width=1\linewidth]{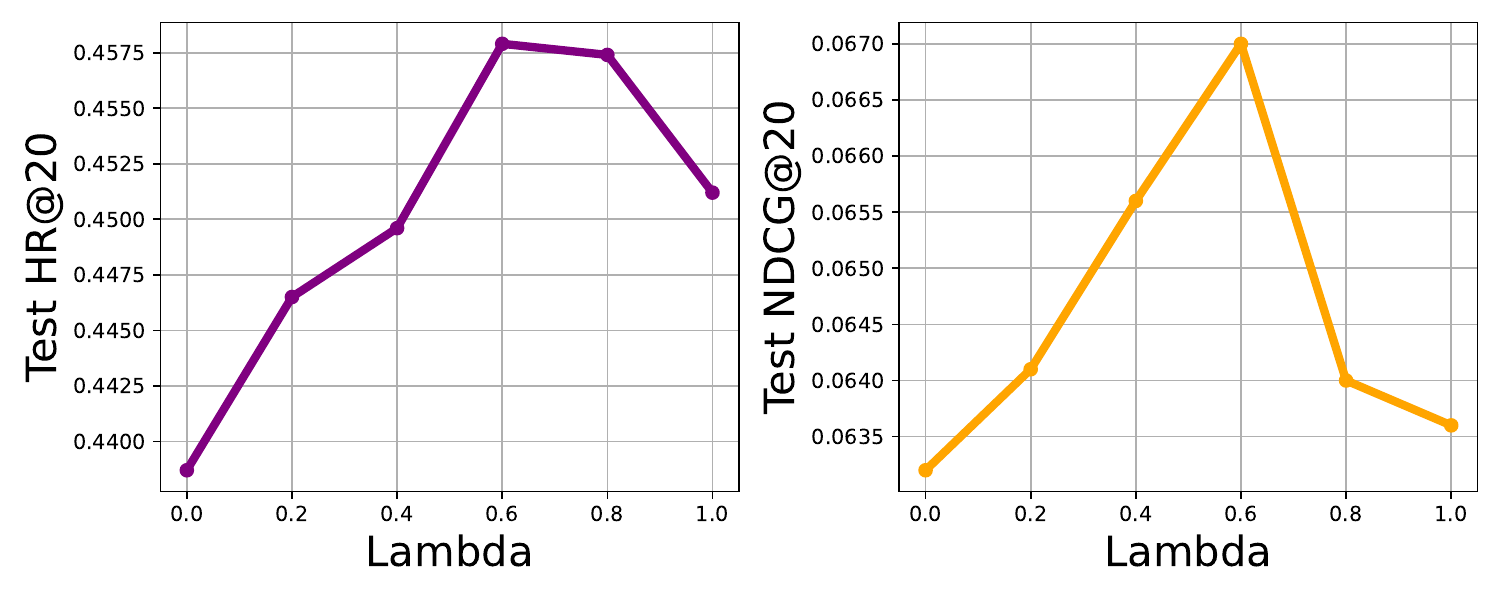}
    \caption{Model performance under different $\lambda$ on MovieLens-1M dataset.}
    \label{fig:hyperparameter2}
\end{figure}

\subsection{Client Selection Bias Analysis (RQ4)}
To investigate the differences in client selection strategies regarding coverage and participation frequency, we report the client selection results of different methods on the Fashion dataset. Specifically, Figure \ref{fig:uniqueclientnumber} presents the number of unique clients selected by each method, while Figure \ref{fig:userheatmap} shows the selection frequency distribution of clients for different methods. From the experimental results, we observe that:
\begin{itemize}
    \item Our client selection method demonstrates a certain degree of selection preference, as revealed by the client selection heatmap. Nevertheless, it achieves higher unique client coverage compared to the random client selection methods and other non-random client selection methods. This result indicates that our method achieves a desirable balance between targeted selection and broad exploration.
    \item Our method selects the largest number of unique clients compared to the random and other baseline methods, demonstrating that our approach enables broader user participation. We attribute this improvement to the staleness-aware reward design, which encourages the exploration of more clients.
    \item The participation frequency under our method exhibits a more strategic pattern than the random method. This indicates that our method not only promotes exploration but also achieves a more balanced selection of clients, ensuring that both active and less frequently selected users are involved in training.
\end{itemize}

\subsection{Hyperparameter Analysis (RQ5)}\label{sec:hy}
To analyze the impact of the proposed algorithm's hyperparameters on model performance, we selected two key hyperparameters and evaluated their effects by assigning different values on MovieLens-1M dataset. Specifically:
First, we investigated how the number of clients selected for training in each federated learning round affects the model's performance by setting $(|\mathcal{U}_{+}| = [50, 100, 200, 500])$.
Then, we examined the impact of $\lambda$, which balances the importance of recommendation accuracy and embedding staleness in the reward function, by setting \(\lambda = [0, 0.2, 0.4, 0.6, 0.8,1]\).
The experimental results are presented in Figure \ref{fig:hyperparameter1}, and \ref{fig:hyperparameter2}. From the results, we observe the following:

\begin{itemize}
    \item As the number of participating clients increases, the recommendation accuracy of the model consistently improves. This is expected because involving more clients in each training round effectively enlarges the training data pool, enabling a greater portion of embeddings to be updated and thus enhancing the overall model performance.

    \item When $\lambda$ is set either too large or too small, the model performance deteriorates significantly. This is reasonable: when $\lambda = 0$, the reward function only includes the staleness term, and the client selector loses its ability to favor clients with higher accuracy contributions. Conversely, when $\lambda = 1$, the reward function completely lacks the staleness guidance, causing the selector to focus excessively on a small subset of high-accuracy clients while neglecting long-tail clients, leading to insufficient updates and degraded model generalization.
\end{itemize}

\vspace{+5em}

\section{Conclusion and Future Work}
\subsection{Conclusion}
In this paper, we first highlight that existing federated recommender systems primarily adopt random client selection strategies, which overlook the issue of statistical heterogeneity in local data, leading to suboptimal model performance. Furthermore, although generic client selection methods in federated learning have shown promising results, directly applying them to recommendation tasks remains challenging due to the high cost of client contribution evaluation and the issue of sparse updates.
To address these challenges, we propose ProxyRL-FRS, a proxy model-guided reinforcement learning framework tailored for client selection in federated recommendation. On one hand, the proposed ProxyNCF branch enables efficient prediction of client contribution without requiring time-consuming local training. On the other hand, the staleness-aware RL agent on the server side leverages a carefully designed reward function to balance recommendation accuracy and embedding freshness, effectively mitigating the sparse update problem. We validate the effectiveness of our approach on three widely used public datasets from different domains and with varying levels of sparsity. Experimental results show that our method consistently outperforms existing state-of-the-art client selection methods in recommendation accuracy while achieving competitive training efficiency.
\subsection{Limitation and Future Work}
One limitation of the current approach is the assumption that users’ local data remains static, making it unsuitable for streaming recommender systems \cite{qu2025efficient,qu2024scalable} where user-item interactions evolve over time. Addressing this limitation in the context of streaming federated recommendation will be the focus of our future work. Moreover, although the proposed method improves both recommendation accuracy and training efficiency, the introduction of the proxy module inevitably incurs additional communication cost. In the future, we plan to explore communication-efficient designs to further reduce overhead while maintaining performance.

\bibliographystyle{IEEEtran}
\bibliography{main}

\vspace{-33pt}

\end{document}